\documentclass[aps,prb,superscriptaddress,twocolumn]{revtex4-2}
\usepackage{graphicx} 
\usepackage[T1]{fontenc}
\usepackage{amssymb}
\usepackage{braket}
\usepackage{physics}
\usepackage{subcaption}
\usepackage{xcolor}
\usepackage{mathtools}
\usepackage{amsmath}
\DeclareMathOperator{\sign}{sign}
\usepackage[scr=rsfs,cal=boondox]{mathalfa}

\usepackage{amsthm}
\usepackage{tikz}
\usepackage{float}
\usepackage{hyperref}
\usepackage[normalem]{ulem}
\usetikzlibrary{quantikz}

\definecolor{jycolor}{rgb}{0.913,0.588,0.478}

\definecolor{jfcolor}{rgb}{0.,0.5,0.5}

\newcommand{\ie}{{\it i.e.}}

\begin{document}

\title{Benchmarking Quantum Convolutional Neural Networks for Classification and Data Compression Tasks}

\author{Jun Yong Khoo}
\affiliation{%
 Institute of High Performance Computing (IHPC), Agency for Science, Technology and Research (A*STAR), 1 Fusionopolis Way, \#16-16 Connexis, Singapore 138632, Republic of Singapore
}
\author{Chee Kwan Gan}
\affiliation{%
 Institute of High Performance Computing (IHPC), Agency for Science, Technology and Research (A*STAR), 1 Fusionopolis Way, \#16-16 Connexis, Singapore 138632, Republic of Singapore
}
\author{Wenjun Ding}
\affiliation{%
 Institute of High Performance Computing (IHPC), Agency for Science, Technology and Research (A*STAR), 1 Fusionopolis Way, \#16-16 Connexis, Singapore 138632, Republic of Singapore
}
\author{Stefano Carrazza}
\affiliation{%
 CERN, Theoretical Physics Department, CH-1211 Geneva 23, Switzerland
}
\affiliation{%
 TIF Lab, Dipartimento di Fisica, Università degli Studi di Milano, Italy
}
\affiliation{%
 INFN, Sezione di Milano, I-20133 Milan, Italy
}
\affiliation{%
 Quantum Research Center, Technology Innovation Institute, Abu Dhabi, UAE
}
\author{Jun Ye}
\affiliation{%
 Institute of High Performance Computing (IHPC), Agency for Science, Technology and Research (A*STAR), 1 Fusionopolis Way, \#16-16 Connexis, Singapore 138632, Republic of Singapore
}
\author{Jian Feng Kong}
\email{kongjf@ihpc.a-star.edu.sg}
\affiliation{%
 Institute of High Performance Computing (IHPC), Agency for Science, Technology and Research (A*STAR), 1 Fusionopolis Way, \#16-16 Connexis, Singapore 138632, Republic of Singapore
}
    
\begin{abstract}
\normalsize
Quantum Convolutional Neural Networks (QCNNs) have emerged as promising models for quantum machine learning tasks, including classification and data compression. This paper investigates the performance of QCNNs in comparison to the hardware-efficient ansatz (HEA) for classifying the phases of quantum ground states of the transverse field Ising model and the XXZ model. Various system sizes, including 4, 8, and 16 qubits, through simulation were examined. Additionally, QCNN and HEA-based autoencoders were implemented to assess their capabilities in compressing quantum states. The results show that QCNN with RY gates can be trained faster due to fewer trainable parameters while matching the performance of HEAs. 
\end{abstract}

\maketitle

\section{Introduction}
Quantum machine learning (QML) leverages quantum computing's principles to enhance classical machine learning algorithms' performance and efficiency. Among various QML models, Quantum Convolutional Neural Networks (QCNNs) have shown significant potential in classification tasks, including both quantum~\cite{Cong2019} and classical data~\cite{Hur2022}.
Inspired by classical convolutional neural networks, QCNNs exploit quantum parallelism and entanglement to process quantum data with short range entanglement structure effectively, and avoids the notorious barren plateau problem by construction due to its logarithmic circuit depth~\cite{Pesah2021}. 

In this study, we seek to demonstrate the extent of its efficiency and advantage over 
the hardware-efficient ansatz (HEA), by comparing their performances on two primary machine learning tasks: phase classification and compression of quantum data. The ground states of transverse field Ising (TFI) model and the XXZ model serve as benchmarks for phase classification, while autoencoder architectures based on QCNN and HEA are evaluated for their compression capabilities of TFI ground states. We also investigate using quantum simulation the impact of different system sizes, \ie{} number of qubits, and classical optimizers on the models' performance.

\section{Methodology}
\subsection{Models}
\textbf{Hardware-Efficient Ansatz} (HEA) is a versatile quantum circuit design that can be adapted to various QML tasks. It typically consists of alternating layers of single-qubit rotations and entangling gates, tailored to the hardware's connectivity. Our preliminary benchmarks showed that for these datasets, the HEA with RY gates obtained similar performances compared to the more general HEA with RX-RZ-RX gates but required only a fraction of the training time. Therefore, we focused on the HEA with RY gates for more detailed benchmark studies against QCNNs.

\textbf{Quantum Convolutional Neural Networks} (QCNNs) extend the concept of classical CNNs to quantum data. They consist of layers of quantum gates that perform convolution and pooling operations on quantum states. Inspired from the competitive performance of HEA with RY gates and together with the observation that these quantum states are real-valued, we hypothesized that real-valued quantum classifiers that are sufficiently expressible would be able to perform the classification task well. To test this hypothesis, we propose and investigate real variants of the original (complex-valued) QCNN with different levels of expressibility and inversely, the number of trainable parameters, leading to trade-offs between performance and training times. 

\subsection{QML Tasks}
\begin{figure*}[ht]
    \centering
    \begin{quantikz}
        \lstick[4]{$\ket{\psi_{\text{test}}}$} & \qw & \gate[4]{U(\theta)} & \meter{}\\
        & \qw & & \qw \\
        & \qw & & \qw \\
        & \qw & & \qw 
    \end{quantikz}
    \hspace{4em}
    \begin{quantikz}
        \lstick[4]{$\ket{\psi_{\text{test}}}$} & \qw & \gate[4]{U(\phi)} & \qw & \gate{\ket{0}} & \qw & \gate[4]{U^\dagger(\phi)} & \qw & \rstick[4]{$\ket{\tilde{\psi}_{\text{test}}}$} \\
        & \qw & & \qw & \gate{\ket{0}} & \qw & & \qw &\\
        & \qw & & \qw \rstick[2]{\hspace{0.5ex}$\ket{\psi_\text{enc}}$} & & \lstick[2]{} & & \qw &\\
        & \qw & & \qw & & & & \qw &
    \end{quantikz}
    \caption{Framework of quantum phase classification (left) and data compression (right) for 4 qubits.}
    \label{Fig.Framework}
\end{figure*}
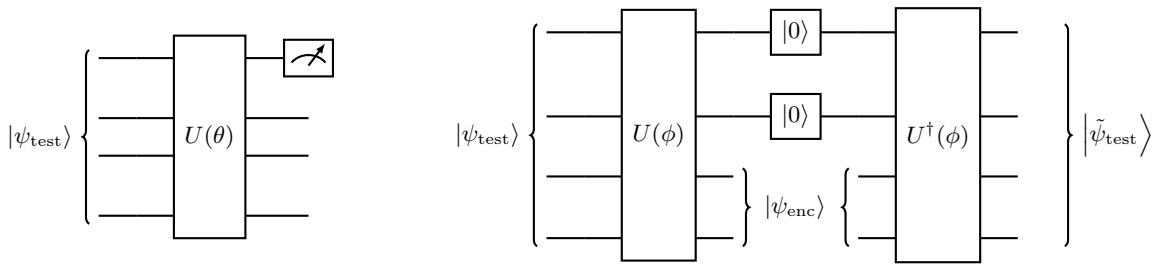

\textbf{Phase Classification}. The phase classification task involves distinguishing between different quantum phases of the ground states of the TFI and XXZ model,
\begin{eqnarray}
    \mathcal{H}_{\rm TFI} &=& -\sum _{j=1}^{N-1}\sigma _j^z \sigma _{j+1} ^z -h \sum _{j=1}^{N} \sigma _j^x, \label{Eq.TFI}\\
    \mathcal{H}_{\rm XXZ} &=& -\sum _{j=1}^{N-1} \left( \sigma_j^x \sigma_{j+1}^x + \sigma_{j}^y \sigma_{j+1}^y + h \sigma_{j}^z \sigma_{j+1}^z\right). \label{Eq.XXZ}
\end{eqnarray}
We adopt the conventional quantum classifier setup, in which the test state $\ket{\psi_{\text{test}}}$ with label $l = \pm 1$, is passed into the trained model $U(\theta)$ with optimized parameters $\theta$. The label predicted by the model $p$, is given by $p = \sign \expval{\psi_{\text{test}}|U^\dagger(\theta)Z_0 U(\theta)|\psi_{\text{test}}}$. We use the mean squared error between $l$ and $p$ over the training dataset as the cost function
(see Fig.~\ref{Fig.Framework}).

\textbf{Data Compression}. For the data compression task, we design quantum autoencoders using QCNN and HEA architectures. These quantum autoencoders~\cite{Romero2017,Bravo2021} aim to compress input quantum states into lower-dimensional representations and then reconstruct the original states with minimal loss. Our setup for each autoencoder involves an encoding phase and a decoding phase. In the former, the test state $\ket{\psi_{\text{test}}}$ is passed into the trained encoder $U(\phi)$ with optimized parameters $\phi$, and finally applying reset gates to the qubits to be discarded to obtain the encoded state $\ket{\psi_\text{enc}}$ of the remaining qubits. In the latter, the $\ket{\psi_\text{enc}}$ is passed with the $\ket{0}$ states of the discarded qubits into the trained decoder $U^\dagger(\phi)$ to obtain the decoded test state  $| \tilde{\psi}_\text{test} \rangle $. The quality of the encoding, can then be determined by the overlap 
$|\langle \tilde{\psi}_{\text{test}}| \psi_{\text{test}} \rangle|^2$
(see Fig.~\ref{Fig.Framework}).

In training of the autoencoder, only the encoder part needs to be trained, as the decoder circuit is the inverse of the encoder. The encoder is trained by minimizing the 
following cost function:
%
\begin{equation}
\label{eqn:enc_cost_fn}
    C = \frac{1}{2} \left(-\sum_{i\in \mathcal{N}_d}Z_i + n_d\right),
\end{equation}
where $\mathcal{N}_d$ and $n_d=|\mathcal{N}_d|$ is the set and number of discarded qubits respectively. Its global minimum occurs when the output state is $\ket{0}^{\otimes n_d}\otimes\ket{\psi_\text{enc}}$, indicating that the discarded qubits can be successfully set to $\ket{0}^{\otimes n_d}$ and $\ket{\psi_\text{test}}$ can be compressed to $\ket{\psi_\text{enc}}$.

\subsection{Simulation Setup}
For the classification task, we conducted experiments with system sizes of 4, 8, and 16 qubits. The training was done using the \texttt{Qibo} library~\cite{Efthymiou2022} on Nvidia A100 GPUs. 
For 4- and 8-qubit systems, we generated labelled datasets of groundstates by performing exact diagonalization to solve the TFI and XXZ models, Eqs.~\eqref{Eq.TFI} and \eqref{Eq.XXZ} respectively, for a range of values of $h$ across their respective phase boundaries. 
For the 16-qubit case, we used optimized variational circuits provided in the \texttt{tensorflow-quantum} repository~\cite{Broughton2021}.

For the data compression task, five input states of 4 and 8 qubits corresponding to the ground states of the TFI model, across the phase boundary of $h=1$, were used. 
QCNN architecture is naturally suited for data compression as each successive layers of QCNN downscales the number of qubits by half, thus $n_d = N (1 - 1/2^l)$, there $l$ is the number of QCNN layers.  For HEA, there are no fixed requirement on downscaling, and hence $n_d$ can be varied from 1 to $N-1$. 
With the trained encoder $U(\tilde{\phi})$, the decoder can be constructed by the inverse of the encoder circuit $U^\dagger(\tilde{\phi})$.

\section{Results}



\begin{figure*}[ht]
\centering
\includegraphics[width=0.9\linewidth]{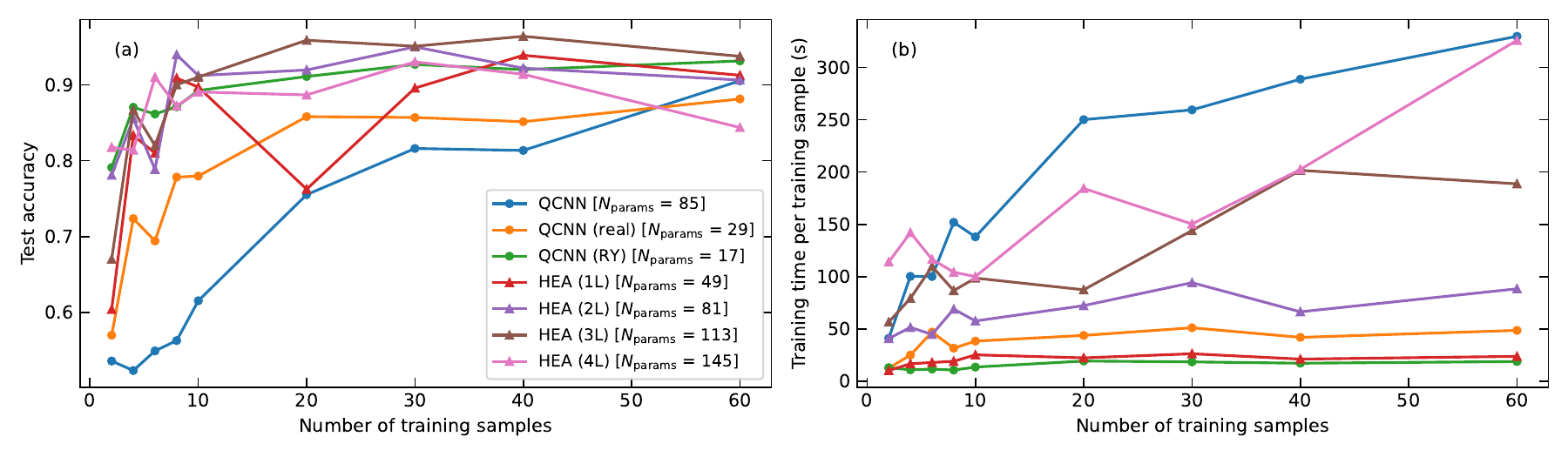}
\caption{
The results of (a) test accuracy for 16 qubits as a function of number of training samples, with their respective training time per training sample shown in (b). In the case of HEA, $n$L denotes $n$ layers. The number of parameters ($N_{\text{params}}$) for each model are shown as well.}
\label{fig:structures}
\end{figure*}

Our results in Fig.~\ref{fig:structures} indicate that QCNN with RY gates showed competitive performance compared to the best performing HEA, with the added benefit of a dramatic reduction in training times due to substantially fewer trainable parameters. For the same number of trainable parameters therefore, QCNN with RY gates is the most efficient architecture for classifying these quantum ground states. The relatively poorer performance of the more expressible QCNN variants can be attributed to trainability issues. HEAs on the other hand, show an increasing performance with expressibility (i.e. number of layers), as well as a corresponding increasing running time per training data (approximately linearly with number of layers). A more detailed analysis was also performed by generating the receiver operating characteristic curve and computing the area under curve.

\begin{table*}[ht]
    \centering
    \begin{center}
    \begin{tabular}{||c c c c||} 
     \hline
     Best Models & Test Accuracy & Training Time / sample & $N_{\text{params}}$ \\ [0.5ex] 
     \hline\hline
     QCNN (RY) & 0.931 & 18.7s & 17 \\ 
     \hline          
     HEA (3L) & 0.938 & 188.7s & 113 \\ [1ex] 
     \hline   
    \end{tabular}
    \end{center}
    \caption{Performance metrics for best QCNN and VQC models for $N_{\text{samples}} = 60$.}
    \label{tab:my_label}
\end{table*}

The autoencoder architectures were assessed on their ability to compress and reconstruct quantum states. Firstly, we note that the generally high reconstruction fidelities imply that all the evaluated models are capable of compressing TFIM ground states to a smaller number of qubits. 
However, QCNNs demonstrated faster training convergence, due to the lower number of trainable parameters in the QCNN models. This demonstrates one key advantage of the QCNN architecture for this task — lower number of trainable parameters leading to faster convergence and lesser trainability issues, with minimal tradeoff in compression capability.

\section{Discussion and Conclusion}
This study highlights the effectiveness of QCNNs, particularly those limited to RY gates, in quantum phase classification and data compression tasks. Consistent across both tasks, we find that QCNNs not only achieve similar performance to HEAs but also benefit from shorter training times due to their simpler structure. The choice of classical optimizer significantly impacts the training efficiency and final model performance. 
Furthermore, during our simulations we observed that Powell is one of the most reliable optimizers for these tasks.
The simulation results obtained will subsequently be compared against implementation on hardware using the \texttt{Qibolab} software library~\cite{Efthymiou2024}.

\bibliographystyle{unsrt}
\bibliography{biblio}
\end{document}